# The impacts of various parameters on learning process and machine learning based performance prediction in online coding competitions

Hardik Patel[1], Purvi Koringa[1]

**Abstract:**

Various parameters affect the performance of students in online coding competitions. Students' behavior, approach, emotions, and problem difficulty levels significantly impact their performance in online coding competitions. We have organized two coding competitions to understand the effects of the above parameters. We have done the online survey at the end of each coding competition, and it contains questions related to the behavior, approach, and emotions of students during online coding competitions. Students are evaluated based on the time and status of the submissions. We have carried out a detailed analysis to address the impact of students' approach, behavior, and emotions on the learning process in online coding competitions. Two difficulty levels are proposed based on the time and status of submissions. The impact of difficulty levels on machine learning-based performance prediction is presented in this research work. Based on time, the coding solution submissions have two classes "Less than 15 minutes" and "More than 15 minutes". There are three classes, "Complete solution", "Partial solution", and "Not submitted at all," based on the submission status. The appropriate approaches are found for both the coding competitions to submit the solution within 15 minutes. Machine learning classifiers are trained and evaluated for the above classification problems. The impacts of mood, emotions, and difficulty levels on the learning process are also assessed by comparing the results of machine learning models for both coding competitions.

**Key Words:** Online coding, emotions, learning, machine learning, problem-solving, and skills

**Software Tools Used:** Scikit learn, Microsoft forms, Microsoft teams, Rapid Miner Studio (academic license),

1. Introduction

The higher education system throughout the world is now focusing on skills rather than traditional assessment systems. The research work proposed in this paper is the first step toward mapping skills to the questions and evaluating different factors that affect students' performance in online coding competitions. Lopez et al. have studied 292 students' learning process strategies and the types of learners by using data from the learning management system and automatic assessment tools [1]. Pereira et al. have deployed a code bench to collect data from introductory programming courses to understand students' behavior during programming activities [2]. Cristobal et al. have presented a review of the prominent publications, the key milestones, the knowledge discovery cycle, and different aspects of education data mining and learning analytics [3]. Integration of education data mining and learning analytics can produce fruitful results for all education stakeholders [3]. Italian Olympiads in Informatics (Olimpiadi Italiane di Informatica - OII) web-based, two distinct platforms are used to perform analytics on programming contest training systems for teachers and students [4]. The evaluation of a student's behavior while developing programming skills and solutions to competitions by using scratch environments [5]. Filvà et al. have proposed an interactive web-based programming platform to collect keyboard and mouse data and classified learners into five affective states: boredom, frustration, distraction, relaxation, and engagement, with an accuracy of 75% [6]. Amigud et al. proposed the integration of learning analytics in the e-learning assessment process to improve academic integrity and assess students' written competitions by a machine learning framework that yields 93% accuracy compared to 12% human accuracy [7]. Ihantola et al. have proposed a literature review and case studies of students' programming processes through different educational data sets and learning analytics techniques [8]. Blikstein et al. have used code snapshots generated by 370 students of introductory undergraduate programming courses. Machine learning models are used on the code snapshots to discover patterns and predict final exam grades [9]. Blikstein et al. have proposed multimodal learning analytics to determine students' learning trajectories and real-time evaluation in offline or online tasks [10]. A review of educational data mining, learning analytics, computer vision applied to assessment, and emotion detection is given in [11] to evaluate student behavior in open-ended programming tasks. In the above literature, learning analytics is applied to understand different factors and parameters which affect students' performance in competitive programming, coding, and academic competitions. Two online coding competitions are organized, and data is collected through survey questions at the end of each online coding competition. The impacts of different parameters like approach, mood, emotions, and difficulty levels on the learning process are evaluated for successful solution development during online coding competitions. The following sections are an analysis of coding competitions, the Application of Machine Learning in online coding, conclusions, and future work.

## 2. Analysis of Coding Competitions

Coding, problem-solving, and logical ability are used to develop the solution to coding problems. Two online coding competitions are organized, and data is collected through survey questions at the end of each coding competition. The first seven survey questions are common in both coding competitions. The last three survey questions are related to students' opinions about coding activities in the first coding competition. However, the previous three survey questions are related to participants' sleep status, mood, and enthusiasm in the second coding competition. 229 and 325 responses were recorded as part of the first and second online coding competitions. The question-wise detailed analysis is described in the following subsection for both the coding completions.

### 2.1 Question wise Analysis of Coding Competitions

1. Have you understood the problem entirely before writing your programming solution?

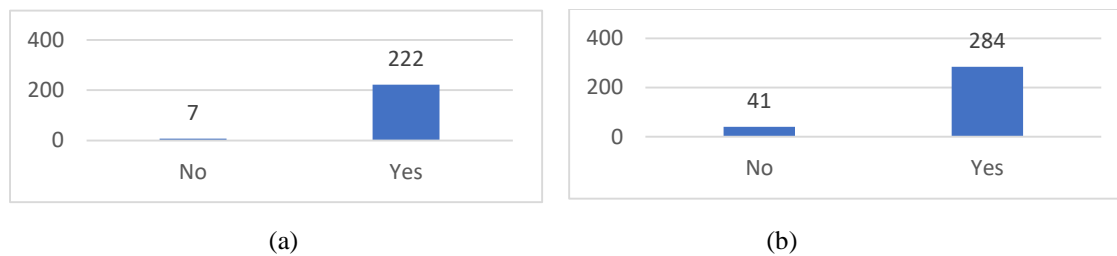

(a)　　　　　　　　　　　　　(b)

Fig.1 Comparison of participants' responses against 1st question (a) Responses recorded after first coding competition (b) Responses recorded after second coding competition.

Fig.1 (a) and (b) show participants' responses to the 1st questions for the first and second coding competitions. The problem of the first coding competition was understood by 222 students (96.94%), and the issue of the dual coding competition was understood by 284 students (87.38%). These responses give hints about the difficulty of both the coding competitions. This question is linked to the understanding skill of students.

2. Have you divided the given problem into sub-problems before writing your code?

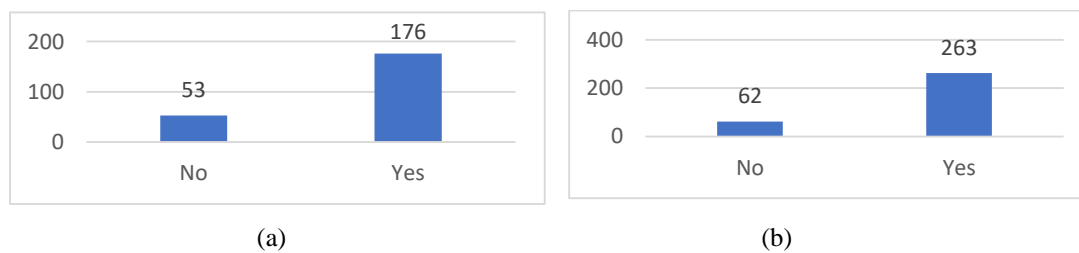

(a)　　　　　　　　　　　　　(b)

Fig.2 Comparison of participants' responses against 2nd question (a) Responses recorded after first coding competition (b) Responses recorded after second coding competition.

Fig.2 (a) and (b) show participants' responses to the 2nd question for the first and second coding competitions. One hundred seventy-six students have divided problems for coding competition one, whereas 263 students have divided problems for coding competition 2. This question is linked to problem-solving skills.

3. Do you need to go through the problem statement again while writing your code?

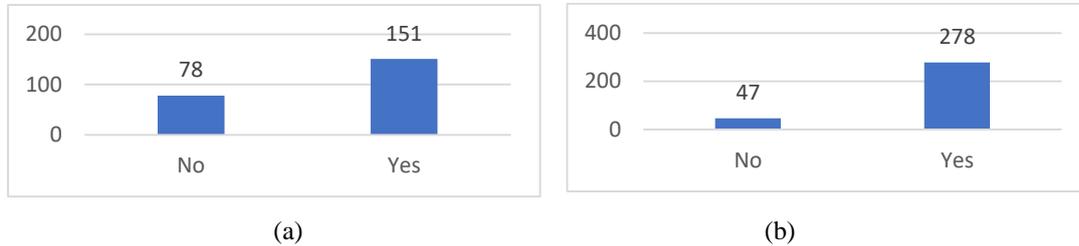

Fig.3 Comparison of participants' responses against 3rd question (a) Responses recorded after first coding competition (b) Responses recorded after second coding competition.

Fig.3 (a) and (b) show participants' responses to 3rd question for the first and second coding competitions, respectively. 78 (34.06%) students have not revisited the problem during the first coding competition, whereas 47 (14.46%) students have not visited the problem during the second coding competition. This question is related to analysis and understanding skills.

4. Have you used the internet while writing your code?

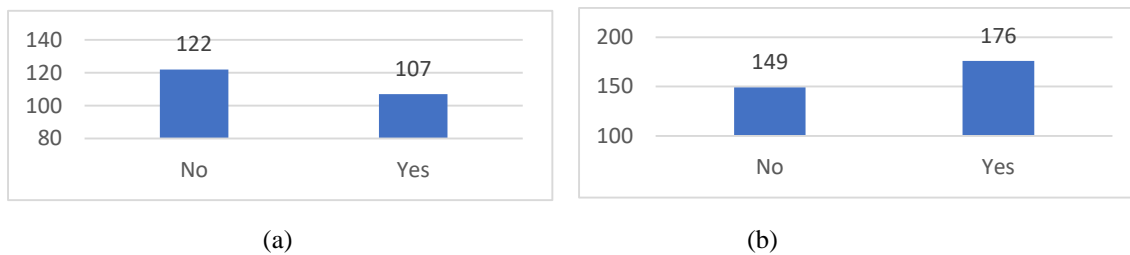

Fig.4 Comparison of participants' responses against 4th question (a) Responses recorded after first coding competition (b) Responses recorded after second coding competition.

Fig.4 (a) and (b) show participants' responses to the 4th question for the first and second coding competitions. Internet was used by 107 (46.72%) students during the first coding competition, whereas 176 (54.15%) students used the internet during the second coding competition. This question is related to logical ability and memory skills.

5. How much time have you taken to complete this program?

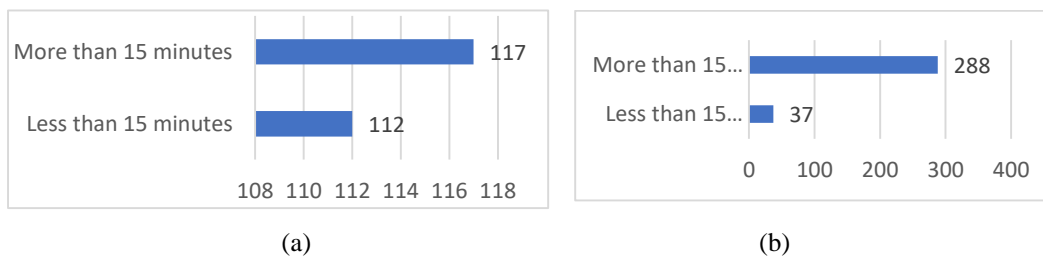

Fig.5 Comparison of participants' responses against 5th question (a) Responses recorded after first coding competition (b) Responses recorded after second coding competition.

Fig.5 (a) and (b) show participants' responses to the 5th question for the first and second coding competitions. 112 (48.90%) students have taken less than 15 minutes to solve the problem of coding competition one, whereas 37 (11.38%) students have taken less than 15 minutes to solve the problem of coding competition 2. This question is related to time management skills.

6. Have you tested your solution by giving different inputs before submission?

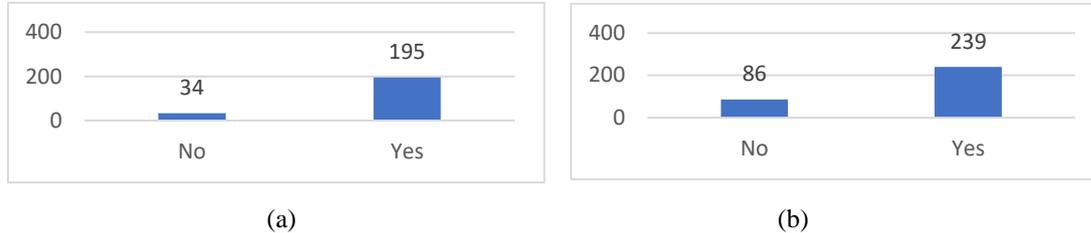

(a)                                              (b)

Fig.6 Comparison of participants' responses against 6th question (a) Responses recorded after first coding competition (b) Responses recorded after second coding competition.

Fig.6 (a) and (b) show participants' responses to the 6th question for the first and second coding competitions. In the first coding competition, one hundred ninety-five students tested their solution before submission. Two hundred thirty-nine students tested their solution before submission in the second coding competition. This question is related to testing and evaluation skills.

7. Have you submitted a complete solution?

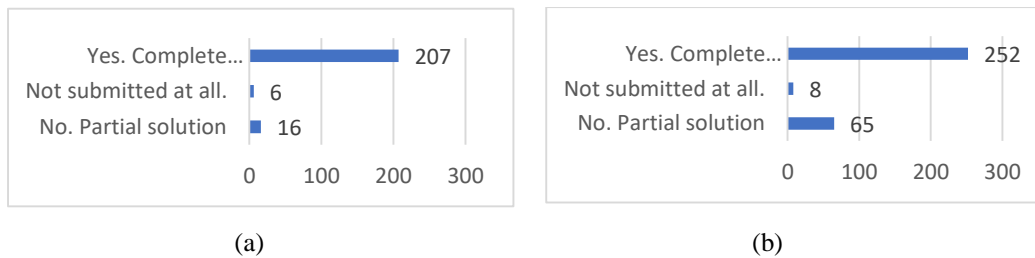

(a)                                              (b)

Fig.7 Comparison of participants' responses against 7th question (a) Responses recorded after first coding competition (b) Responses recorded after second coding competition.

Fig.7 (a) and (b) show participants' responses to the 7th question for the first and second coding competitions. 207 (90.39%) students have submitted complete solutions for coding competition 1, whereas 252 (77.53%) students have submitted complete solutions for coding competition 2. This question is related to task completion or deadline management skills.

8. How was your mood when you started this activity?

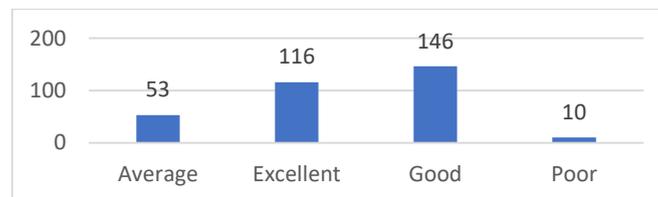

Fig.8. Responses were recorded for the 8th question after the second coding competition.

Fig.8 shows the participant's responses to the 8th question for the second coding competition. 116 (35.69%) students had an excellent mood, whereas 146 (44.92%) students had a good mood during the second coding competition. This question is related to the status of the mind and brain during coding activity (emotional level).

9. Have you taken enough sleep and felt good before this programming activity?

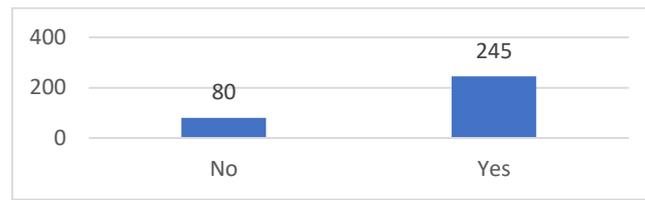

Fig.9. Responses were recorded for the 9th question after the second coding competition.

Fig.9 shows the participant's responses to the 9th question for the second coding competition. 245 (75.38%) students have taken enough sleep and felt good during the second coding competition. This question is related to the relaxation of the mind and brain during coding activity (physical and mental status).

10. Have you completed this activity with full enthusiasm?

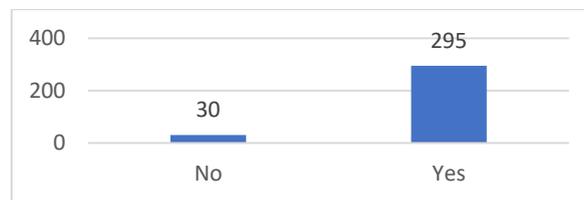

Fig.10. Responses were recorded for the 10th question after the second coding competition.

Fig.10 shows participants' responses to the 10th question for the second coding competition. 295 (90.77%) students completed this activity with full enthusiasm during the second coding competition. This question relates to feeling good about coding activity (Student's preference about coding activity).

## 2.2 Overall Analysis of Coding competitions

There are two essential criteria to determine the successful completion of the activity. Students have learned new skills through these coding competitions if they have submitted either a complete solution or a partial solution. The difficulty perceived by the student can be measured through the time taken by students to complete both the coding competitions and the status of submission (full or partial or not submitted). The first coding competition was completed by 112 students within 15 minutes, whereas only 37 students completed the second coding competition within 15 minutes. The second coding competition problem was difficult for students as compared to the first coding competition problem. Approximately 90% of students have submitted a complete solution for the first coding competition, whereas about 82% of students have submitted a complete solution for the second coding competition. Answers to questions 5 and 7 are pretty helpful in determining the comparative difficulty of both coding competition problems. Through learning analytics, we tried to identify which factors are essential to answer coding competition problems in less than 15 minutes. To determine the weightage of different aspects to respond to the first coding problem within 15 minutes, refer to table 1.1 given below. Out of these 112 students who have submitted the solution successfully, all (100%) have understood the problem. 22 (20%) students have not divided the problem into mini-goals, whereas 90 (80%) students have divided the problem statement. 52 (46%) students have revisited the problem statement, and 60 (54%) students have not revisited the problem. 79 (70%) students have not used the internet, whereas 33 (30%) students have used the internet to solve a problem. 105 (94%) students have tested their solution, whereas 7 (6%) students have not tested it.

Table 1.1 Analysis of 112 students who have submitted the first coding competition solution within 15 minutes.

| **Understood** | **Divide** | **Revisit** | **Internet used** | **Testing** | **Submission** |
|---|---|---|---|---|---|
| Yes-112 | Yes-90 | Yes-52 | Yes-33 | Yes-105 | Complete Solution-112 |
| No-0 | No-22 | No-60 | No-79 | No-7 | Partial solution-0 |
|  |  |  |  |  | Not submitted at all-0. |

Thirty-seven students submitted a second coding competition solution within 15 minutes. Out of 37 students who submitted a second coding competition solution within 15 minutes, 35 (95%) students understood the problem, whereas 2 (5%) students did not understand it. 34 (92%) students have divided the problem into mini-goals, whereas 3 (8%) students have not divided the problem, 14 (38%) students have revisited the problem statement whereas, and 23 (62%) students have not revisited the problem statement, 8 (22%) students have used internet whereas, 29 (78%) have not used the internet, 33 (89%) have tested the solution whereas, 4 (11%) have not tested the solution.

Table 1.2 Analysis of 37 students who have submitted a second coding competition solution within 15 minutes.

| Understood | Divide | Revisit | Internet used | Testing | Submission | Mood | Sufficient Sleep | Enthusiasm |
|---|---|---|---|---|---|---|---|---|
| Yes-35 | Yes-34 | Yes-14 | Yes-8 | Yes-33 | Complete Solution-34 | Excellent-27 | Yes-31 | Yes-37 |
| No-2 | No-3 | No-23 | No-29 | No-4 | Partial Solution-3 | Good-10 | No-6 | No-0 |
|  |  |  |  |  | Not submitted at all-0 | Average-0 |  |  |
|  |  |  |  |  |  | Poor-0 |  |  |

34 (92%) students have Submitted complete solutions, whereas 3 (8%) have submitted partial solutions, 27 (73%) students have excellent mood, whereas 10 (27%) have a good mood. 31 (84%) students have taken sufficient sleep whereas, 6 (16%) have not taken enough sleep, all 37 (100%) students were enthusiastic.

**2.3 Difficulty Level of Coding Problems and Performance Measurement**

Coding problems may be categorized as simple, moderate, difficult, and very difficult for students. The difficulty level of coding problems can be predetermined by using intuition. In this research work, the difficulty level is determined through the submissions and responses of students. There are two difficulty level measurements based on the time and status of the submission. The complete solution was submitted by 90.39% of students in the first coding problem. The complete solution was submitted by 77.53% of students in the second coding problem. The relative difficulty level (DL) involved in these problems is given by equation 1.

$$DL = 1 - (\text{percentage of students who have submitted complete solution}) \qquad (1)$$

According to equation (1), the difficulty level of the first coding problem is around 0.0961 (9.61%), whereas the difficulty level of the second coding problem is around 0.2247 (22.47%). The difficulty level can also be measured by using the time to submit the solution. 112 (48.90%) students solved the first coding problem within 15 minutes, whereas 37 (11.38%) students solved the second coding problem within 15 minutes. The second definition of difficulty level is given by equation (2).

$$DLT = 1 - (\text{percentage of students who have submitted solution in less than 15 minutes}) \quad (2)$$

Where DLT is a difficulty level concerning time, according to this definition, the difficulty level of the first problem is around 0.511 (51.1%). In contrast, the difficulty level of the second coding problem is around 0.8862 (88.62%). Difficulty level measurement of the coding problem is a relative measure, and it is up to the faculty, university, and teaching community. Above two types of difficulty levels are proposed to show that there is a possibility to present different kinds of difficulty levels according to the parameter under investigation or skills to be measured. Performance measurement can be done by using proposed difficulty levels.

Table 1.3 Performance of students in the first coding competition

| Performance | No. of Students (%) |
|---|---|
| Excellent | 112 (48.90%) |
| Good | 95 (41.48%) |
| Average | 16 (6.99%) |
| Poor | 6 (2.62%) |

Students who have submitted a complete solution within 15 minutes can be awarded excellent performance, students who have submitted a complete solution between 15 minutes to 1 hour can be awarded good performance, students who have partially submitted a solution can be awarded average performance, and students who have not submitted solution can be awarded poor performance. According to table 1.3, almost 90% of students have excellent and good performance in the first coding competition. Approximately 49% of students have excellent performance in the first coding competition.

Table 1.4 Performance of students in the second coding competition

| Performance | No. of Students (%) |
|---|---|
| Excellent | 37 (11.38%) |
| Good | 215 (66.15%) |
| Average | 65 (20%) |
| Poor | 8 (2.46%) |

According to table 1.4, only 11.38% of students have excellent performance in the second coding competition, which indicates the difficulty level of the coding problem. Almost 86% of students had a good and average performance in the second coding competition.

## 3. Application of Machine Learning in online coding

Different machine learning algorithms are applied to the data collected to generate insights about factors affecting students' performance in coding competitions. Rapid Miner studio academic version is used to apply machine learning in this research work.

Table 1.5 Performance comparison of different classification algorithms to classify the time required to complete the competition (less than 15 minutes, more than 15 minutes) for the first coding competition

| Model | F1 Measure | Standard Deviation | Gains | Total Time | Training Time (1,000 Rows) | Scoring Time (1,000 Rows) |
|---|---|---|---|---|---|---|
| Naive Bayes | 0.7216 | 0.0708 | 22.0 | 1651.0 | 161.6 | 217.4 |
| Generalized Linear Model | 0.6461 | 0.0856 | 14.0 | 1181.0 | 262.0 | 173.9 |
| Logistic Regression | 0.7083 | 0.0526 | 20.0 | 946.0 | 266.4 | 141.3 |
| Fast Large Margin | 0.5350 | 0.1460 | 8.0 | 1727.0 | 144.1 | 119.6 |
| Deep Learning | 0.6528 | 0.0411 | 16.0 | 1945.0 | 1358.1 | 217.4 |
| Decision Tree | 0.6483 | 0.0787 | 6.0 | 981.0 | 117.9 | 119.6 |
| Random Forest | 0.7050 | 0.0594 | 20.0 | 12516.0 | 165.9 | 1413.0 |
| Gradient Boosted Trees | 0.6302 | 0.0520 | 10.0 | 11676.0 | 554.6 | 152.2 |
| Support Vector Machine | 0.6023 | 0.1204 | 18.0 | 2922.0 | 777.3 | 228.3 |

Table 1.6 Performance comparison of different classification algorithms to classify submission status for a first coding competition

| Model | Accuracy (%) | Standard Deviation (%) | Gains | Total Time (sec) | Training Time (1000 Rows) | Scoring Time (1,000 rows) |
|---|---|---|---|---|---|---|
| Naive Bayes | 78.46 | 6.44 | -16.0 | 1186.0 | 196.5 | 120.9 |
| Generalized Linear Model | 69.23 | 7.69 | -30.0 | 1086.0 | 414.8 | 263.7 |
| Logistic Regression | 83.08 | 8.43 | -12.0 | 1997.0 | 406.1 | 285.7 |
| Fast Large Margin | 83.08 | 6.44 | -8.0 | 3666.0 | 139.7 | 472.5 |
| Deep Learning | 90.77 | 3.44 | 0.0 | 1813.0 | 724.9 | 285.7 |
| Decision Tree | 90.77 | 3.44 | 0.0 | 1286.0 | 170.3 | 120.9 |
| Random Forest | 78.46 | 6.44 | -16.0 | 3154.0 | 139.7 | 351.6 |
| Gradient Boosted Trees | 90.77 | 3.44 | 0.0 | 8895.0 | 449.8 | 219.8 |
| Support Vector Machine | 90.77 | 3.44 | 0.0 | 3057.0 | 174.7 | 307.7 |

Time to complete the coding competition can be used to check students' time management skills. It can also measure the relative difficulty levels of coding problems. Students have completed the coding problem in less than 15 minutes or more than 15 minutes, and it is modeled as a binary classification problem for both the coding competitions. Data related to sleep, mood, and enthusiasm were collected during a second coding competition. It is used during submission time-based classification, resulting in better performance for all classifiers. Submission status classification is not improved due to sleep, mood, and enthusiasm-related features. Submission status classification largely depends on the difficulty level or complexity of the coding competitions. Table 1.5 shows the performance comparison of different classification algorithms to classify the time taken to solve the first coding competition. F1 measure shows that the performance of Naïve Bayes is best among all the algorithms. Logistic regression and random forest algorithms are performing well, in second place. Table 1.6 shows the performance comparison of different machine learning algorithms to classify the submission status for the first coding competition. The best submission status-based classification performance is based on deep learning, decision trees, gradient boosted trees, and support vector machines.

Table 1.7 Performance comparison of different classification algorithms to classify the time taken to complete the competition (less than 15 minutes, more than 15 minutes) for the second coding competition

| Model | F Measure | Standard Deviation | Gains | Total Time | Training Time (1,000 Rows) | Scoring Time (1,000 Rows) |
|---|---|---|---|---|---|---|
| Naive Bayes | 0.9413 | 0.0400 | 0.0000 | 847.0 | 70.8 | 100.0 |
| Generalized Linear Model | 0.9413 | 0.0400 | 0.0000 | 936.0 | 116.9 | 84.6 |
| Logistic Regression | 0.9413 | 0.0400 | 0.0000 | 966.0 | 166.2 | 84.6 |
| Fast Large Margin | 0.9460 | 0.0277 | 2.0000 | 2088.0 | 230.8 | 100.0 |
| Deep Learning | 0.9413 | 0.0400 | 0.0000 | 2322.0 | 732.3 | 223.1 |
| Decision Tree | 0.9413 | 0.0400 | 0.0000 | 1391.0 | 86.2 | 100.0 |
| Random Forest | 0.9413 | 0.0400 | 0.0000 | 3398.0 | 70.8 | 253.8 |
| Gradient Boosted Trees | 0.9514 | 0.0298 | 4.0000 | 14243.0 | 987.7 | 200.0 |
| Support Vector Machine | 0.9520 | 0.0296 | 4.0000 | 3016.0 | 110.8 | 153.8 |

Table 1.8 Performance comparison of different classification algorithms to classify submission status for a second coding competition

| Model | Accuracy (%) | Standard Deviation | Gains | Total Time | Training Time (1,000 Rows) | Scoring Time (1,000 Rows) |
|---|---|---|---|---|---|---|
| Naive Bayes | 82.69 | 0.0969 | 16.0 | 1449.0 | 135.4 | 153.8 |
| Generalized Linear Model | 78.13 | 0.0898 | 6.0 | 1738.0 | 353.8 | 169.2 |
| Logistic Regression | 83.80 | 0.0930 | 18.0 | 3059.0 | 344.6 | 376.9 |
| Fast Large Margin | 82.75 | 0.1024 | 16.0 | 6117.0 | 341.5 | 492.3 |
| Deep Learning | 82.75 | 0.0703 | 16.0 | 2657.0 | 609.2 | 246.2 |
| Decision Tree | 87.02 | 0.0637 | 20.0 | 1939.0 | 107.7 | 207.7 |
| Random Forest | 83.86 | 0.0543 | 14.0 | 5577.0 | 110.8 | 507.7 |
| Gradient Boosted Trees | 84.91 | 0.0706 | 16.0 | 11069.0 | 341.5 | 384.6 |
| Support Vector Machine | 82.46 | 0.0741 | 14.0 | 5972.0 | 190.8 | 500.0 |

Table 1.7 shows the performance comparison of different machine learning models to classify submissions status for the second coding competition. Support vector machine and gradient boosted trees have the best performance in predicting the time taken to complete the second coding competition. The inclusion of mood and emotion-related questions has improved the performance of machine learning algorithms for the second coding competition. If table 1.5 and table 1.7 are compared, it clearly shows better performance for the second coding problem due to the inclusion of mood and emotion-related questions. Table1.8 shows the performance comparison of different classification algorithms to classify submission status for the second coding competition. The decision tree has the best performance compared to other algorithms for the submission status-based classification during the second coding competition. If table 1.6 is compared with table 1.8, it clearly shows that the performance of machine learning algorithms is good for the submission status-based classification during the first coding competition. The inclusion of mood and emotional-related questions has improved the overall performance of all algorithms. However, the performance of machine learning algorithms is still good in predicting submission status for the first coding competition. Prediction of submission status is not improved remarkably due to the inclusion of emotions and mood-related questions because the second coding problem is more challenging than the first one. Here difficulty plays a role in predicting the submission status of both the coding competitions.

## 4. Conclusions and Discussion

Overall statistical analysis shows promise if students' behavior, approach, and emotions are recorded automatically through interactive web technology and multimodal learning analytics. The first step is to understand the learning process while solving coding problems. The analysis clearly shows that during the first coding competition, students who have understood the problem, divided the problem, not revisited the problem, not used the internet, and tested the solution have the highest possibility of solving the coding problem in less than 15 minutes. The performance of Naïve Bayes is best among all the algorithms for the submission time-based classification (less than 15 minutes, more than 15 minutes) in the first coding competition. Deep learning, decision trees, gradient boosted trees, and support vector machines have the best performance for the submission status-based classification in the first coding competition. In the second coding competition, the support vector machine and gradient boosted trees had the best performance for the submission time-based classification (less than 15 minutes, more than 15 minutes). The decision tree has the best performance compared to other algorithms for the submission status-based classification in the second coding competition. The better performance of machine learning classifiers is achieved for the second coding problem due to the inclusion of mood and emotions related questions. The performance of machine learning algorithms is suitable for submission status-based classification in the first coding competition. The first coding problem was not as complex as the second one, which also plays a significant role in predicting submission status. The analysis clearly shows that students who have submitted solutions in less than 15 minutes have either excellent or good moods. Most students who have submitted complete solutions also have either excellent or good moods. The inclusion of mood and emotions related questions can improve overall performance, but difficulty level matters in the prediction of submission status.

## 5. Future work

This research work is conducted based on survey questions. It is the first step towards understanding the impacts of approach, mood, emotions, and difficulty level on the learning process while solving the coding problems. Students' behavior, strategy, and emotions will be understood through automated and interactive web technology and multimodal learning analytics in future work. The terms scorecard or mark sheet or transcript, or grade card are widely used worldwide. There should be a concept of skill cards after every semester so that students can analyze their skills. Mapping questions to skills is crucial for generating skill cards for students. Skill cards can be caused by evaluating the performance of students for skills. Coding competitions or aptitude questions are also helpful in developing skill cards for students. Universities can have one dedicated office to implement skill card generation and distribution policies. Learning can be personalized by giving skill cards to students so they can assess their skills and come to know about their strengths and weakness in terms of skills. Students have a clear idea about their performance in particular skills to make strategies to improve or retain their performance in specific skills. Students can learn about skills to improve through skill cards. In contrast, universities can generate substantial data sets, and learning analytics can be applied to the data set generated through skill cards. The concept of skill cards can improve the overall performance of students and universities. The research work and future work proposed here are helpful to all the higher education stakeholders.